\documentclass
[preprint,amsmath,amssymb,a4paper,superscriptaddress]
{revtex4}
\usepackage{graphicx}
\usepackage{here}
\begin{document}

\title{Interface reconstruction in V-oxide heterostructures
determined by x-ray absorption spectroscopy}

\author{H. Wadati}
\email{wadati@phas.ubc.ca}
\affiliation{Department of Physics and Astronomy, 
University of British Columbia, Vancouver, 
British Columbia V6T 1Z1, Canada}

\author{D. G. Hawthorn}
\affiliation{Department of Physics and Astronomy, 
University of British Columbia, Vancouver, 
British Columbia V6T 1Z1, Canada}

\author{J. Geck}
\affiliation{Department of Physics and Astronomy, 
University of British Columbia, Vancouver, 
British Columbia V6T 1Z1, Canada}

\author{T.~Z.~Regier}
\affiliation{Canadian Light Source, University of Saskatchewan, 
Saskatoon, Saskatchewan S7N 0X4, Canada}

\author{R.~I.~R.~Blyth}
\affiliation{Canadian Light Source, University of Saskatchewan, 
Saskatoon, Saskatchewan S7N 0X4, Canada}

\author{T.~Higuchi}
\affiliation{Department of Advanced Materials Science, 
University of Tokyo, Kashiwa, Chiba 277-8561, Japan}

\author{Y. Hotta}
\affiliation{Department of Advanced Materials Science, 
University of Tokyo, Kashiwa, Chiba 277-8561, Japan}

\author{Y. Hikita}
\affiliation{Department of Advanced Materials Science, 
University of Tokyo, Kashiwa, Chiba 277-8561, Japan}

\author{H. Y. Hwang}
\affiliation{Department of Advanced Materials Science, 
University of Tokyo, Kashiwa, Chiba 277-8561, Japan}
\affiliation{Japan Science and Technology Agency, 
Kawaguchi, 332-0012, Japan}

\author{G. A. Sawatzky}
\affiliation{Department of Physics and Astronomy, 
University of British Columbia, Vancouver, 
British Columbia V6T 1Z1, Canada}

\date{\today}
\begin{abstract}
We present an x-ray absorption study of the dependence 
of the V oxidation state on the thickness of 
LaVO$_3$ (LVO) and capping LaAlO$_3$ (LAO) layers 
in the multilayer 
structure of LVO sandwiched between LAO. 
We found that the change of the valence of V 
as a function of LAO layer thickness 
can be qualitatively explained by a transition 
between electronically reconstructed interfaces 
and a chemical reconstruction. 
The change as a function of LVO layer thickness 
is complicated by the presence of a considerable 
amount of V$^{4+}$ in the bulk of 
the thicker LVO layers. 
\end{abstract}
\maketitle
The interfaces of hetero-junctions composed 
of transition-metal oxides have recently 
attracted great interest. For example, 
the interface between two band 
insulators SrTiO$_3$ (STO) and LaAlO$_3$ (LAO) 
is especially interesting due to its 
metallic \cite{Hwang2} and even 
superconducting properties \cite{superLAOSTO}. 
The interface between a band insulator 
STO and a Mott 
insulator LaTiO$_3$ (LTO) also exhibits metallic 
conductivity \cite{HwangTi,shibuyaLTOSTO,okamotoLTOSTO}. 
Recently Takizawa {\it et al.} \cite{takizawaLTOSTO} 
reported photoemission spectra of this interface 
with a clear Fermi cut-off, indicating 
a metallic interface layer. 
In this paper, we investigate the electronic structure 
of multilayers consisting of a band insulator 
LAO and a Mott insulator LaVO$_3$ (LVO). 
LAO is a band insulator with a large 
band gap of about 5.6 eV \cite{LAO}, 
whereas LVO is 
a Mott-Hubbard insulator with a Mott gap of 
about 1.0 eV \cite{opt1}. 
One striking characteristic of this interface 
is the absence of a 
polar discontinuity because both LAO and LVO 
consist of alternating stacks of polar layers 
(LaO$^+$ and AlO$_2^-$ or VO$_2^-$). 
Photoemission study of this interface has been reported 
in Refs.~\cite{Hotta, WadatiLAOLVO, takizawacond}. 
From the V $2p$ core-level 
spectra measured by soft x-ray \cite{Hotta} and 
hard x-ray \cite{WadatiLAOLVO, takizawacond}, the valence of V
in LVO was found to be partially converted 
from V$^{3+}$ to V$^{4+}$ at the interface. 
Such a hole doping of LVO 
at the interface is expected to 
cause metallicity in this interface, but 
from the valence-band spectra \cite{WadatiLAOLVO}, 
a Mott-Hubbard gap of LVO was found to remain 
open at the interface, indicating an insulating 
nature of this interface. 
This is consistent with transport studies showing 
increasing hole doping and conductivity with decreasing 
LAO thickness, but with an insulating temperature 
dependence \cite{Higuchicond}. 
The absence of metallicity 
is in contrast to the case of LAO/STO 
and LTO/STO, where electron doping of STO 
causes metallic conductivity 
at orders of magnitude lower carrier 
density than in LVO. 

In this study we investigated 
the electronic structure of V oxide thin films 
by x-ray absorption spectroscopy (XAS). 
Regarding the determination of the valence of V, 
XAS has several advantages over photoemission 
spectroscopy (PES). XAS is rather bulk-sensitive 
with the average probing depth of $>50$ $\mbox{\AA}$ 
in total-electron-yield (TEY) mode and 
$>200$ $\mbox{\AA}$ 
in fluorescence-yield (FY) mode, 
whereas that of PES is 
at most $\sim 50$ $\mbox{\AA}$ even in the 
hard x-ray region. In the 
V $2p \rightarrow 3d$ XAS, a core hole is 
screened by an additional $3d$ electron, so XAS 
is more sensitive to the multiplet structure, 
and thus to the valence of V, than PES. XAS can be 
also applied to insulating materials due 
to its charge-neutral process. On the other 
hand, PES has a large advantage of detecting 
the valence-change-induced chemical shift 
and observing the structure near the Fermi level 
thus determining whether the system is metallic 
or insulating. 

From the XAS studies 
of LAO-sandwiched LVO thin films, 
we found that the V $2p$ XAS spectra depend on both the 
thickness of LVO and the thickness of the 
LAO capping layers. 
The dependence on LAO thickness can be qualitatively 
explained by the transition from an 
``electronic reconstruction'' at the LAO/LVO interface 
to a ``chemical reconstruction'' involving 
surface off-stoichiometry \cite{foot1}, but 
that on LVO thickness indicates a more complex evolution. 
 
The V oxide thin films were fabricated  
using the pulsed laser deposition (PLD) technique. 
The details of the sample fabrication and characterization 
are described in Refs.~\cite{HottaLVO, Hotta, Higuchicond}. 
The characterization 
of the electronic structure of uncapped LVO thin films 
by x-ray photoemission spectroscopy is described 
in Ref.~\cite{WadatiuncapLVO}. The 
films were grown on the TiO$_2$-terminated STO or 
AlO$_2$-terminated LAO substrates at an oxygen 
pressure of 10$^{-6}$ Torr, 
except for LaVO$_4$ (10$^{-2}$ Torr), 
using a KrF excimer laser 
($\lambda = 248$ nm). 
Schematic views of the present thin films 
are shown in Fig.~1. 
Sample (a) was 50 uc LaVO$_4$ on STO without any 
capping layers as a reference for V$^{5+}$. 
Sample (b) was 50 uc SrVO$_3$ on STO without any 
capping layers as a reference for V$^{4+}$. 
Sample (c) consisted of $y$ uc LVO sandwiched by $x$ uc 
LAO and the LAO substrate. 
X-ray absorption experiments were 
performed at 11ID-1 (SGM) of the Canadian Light Source. 
The spectra were measured both in TEY and FY modes. 
As for the spectra we show in this paper, 
there was almost no difference between these two modes, 
so we show only the spectra measured 
in the TEY mode. 
Changing the exit slit width, which changes the experimental 
resolution from 90 to 800 meV, did not have an appreciable 
effect on the lineshape. Therefore, 
the total energy resolution was set to 800 meV 
in order to maximize intensities 
without sacrificing spectral details. 
All the spectra were measured at room temperature. 

Figure 2 (a) shows the V $2p$ XAS spectrum of 50 uc LaVO$_4$ 
thin films on STO(001). 
The spectrum displayed sharp features 
characteristic of V$^{5+}$ ($d^0$) oxides \cite{GermanG}. 
Figure 2 (b) shows the V $2p$ XAS spectrum of 50 uc SrVO$_3$ 
thin films on STO(001) with features characteristic of V$^{4+}$. 
Figure 2 (d) displays 
the mixed valent (V$^{3+}$ and V$^{4+}$) character 
in the V $2p$ XAS spectra of LAO-sandwiched 
LVO thin films, where linear backgrounds have been 
subtracted as shown in Fig.~2 (c). 
The V $2p$ XAS spectra depend on 
the thickness of both LVO and LAO capping layers. 
By comparing these spectra with that of 
YVO$_3$ \cite{Penn}, the most sensitive region 
displaying V$^{4+}$ character occurs around 520 eV, 
marked by an arrow in Fig.~2 (d). From this we note 
that even the (8, 3) sample contains a small 
amount of V$^{4+}$. To obtain a V$^{3+}$ 
characteristic spectrum, the V$^{4+}$ component 
(the spectrum of SVO (50 uc)/STO), 
which amounts to 5 \% in this case, was 
subtracted from the spectrum of (8, 3). 

By using this V$^{3+}$ spectrum and 
the V $2p$ XAS spectrum of SVO (50 uc)/STO 
as V$^{3+}$ and V$^{4+}$ references, respectively, 
we can determine the average 
valence of V in the other samples. 
Figure 3 shows this determination of the valence of 
V by using a linear combination of V$^{3+}$ and V$^{4+}$. 
The valence of V in (8, 3) is 3.05 
as discussed above, and those of 
(3, 3), (8, 20), and (3, 20) were 
determined to be 3.45, 3.2, and 3.3 
($\pm 0.1$), respectively. 

In order to understand these results, we studied 
the following models. 
In the present samples, 
both the LAO and LVO layers are polar, and 
consist of alternating stacks of 
LaO$^+$ and AlO$_2^-$ or VO$_2^-$ layers. 
As recently discussed 
by Nakagawa {\it et al.} \cite{nakagawa}, 
electronic reconstruction \cite{hesperrecon} 
occurs at the interface of 
polar layers and nonpolar layers 
in order to prevent the divergence of 
Madelung potential, i.e., 
the so-called polar catastrophe \cite{catas}. 
When the V valence changes 
in the topmost LVO layer, 
the divergence of the electrostatic potential 
disappears in the LVO layers, 
but a dipole shift occurs in the LAO layers. 
This type of reconstruction is called 
``electronic reconstruction'' as shown in Fig.~4 (b). 
This model assumes that the electronic reconstruction 
causes mixed valence of V only at one interface. 
In another case when surface off-stoichiometry 
(for example, O vacancies) occurs in the topmost 
LAO layer and part of the excess electrons 
created at the same time are transferred to the 
opposite side of the polar layer, 
the divergence of the electrostatic potential 
disappears as shown in Fig.~4 (c) \cite{footf}. 
In both cases, we can obtain the valence of V expected to 
be observed in XAS measurements. 
XAS measures a weighted average of V valence 
over the whole sample. 
The weighting is determined by the 
exponential decay of the secondary electron yield 
($\propto \exp (-z/\lambda_e)$ from 
layers at a distance $z$ from the outer surface). 
We present the results of this model calculation 
using the probing depths $\lambda_e$ of 
50, 200, and $\infty$ $\mbox{\AA}$ in Fig.~4 (a), 
showing only a weak dependence on the choice of $\lambda_e$. 
In experiment, the valence of V decreases 
with increasing LAO thickness, which is qualitatively 
consistent with the transition from electronically 
reconstructed interfaces to a chemical reconstruction to avoid 
the potential divergence in LAO layers. 
However, the change of the valence of V 
as a function of LVO layers is not systematic. 
The observed average valence is $3.2+$ or $3.3+$ in the 
(8, 20) and (3, 20) samples, respectively. The existence of 
such a large amount of V$^{4+}$ cannot be explained by 
changing the valence of V from $3+$ to $4+$ 
only at the interface and suggests off-stoichiometry effects 
in the bulk of thick LVO layers. Such a change of 
the valence of V can be caused by excess O or 
deficiencies of La or V, and also 
influences the effects of the 
polar catastrophe in LVO layers. 

In this study, we succeeded in determining the valence of V in 
multilayers composed of a band insulator LAO and 
a Mott insulator LVO by means of 
x-ray absorption spectroscopy. 
Here we made use of its sensitivity 
to bulk states and to the multiplet structure. 
We found that the V $2p$ XAS spectra depend on the 
thickness of both LVO and LAO capping layers. 
The dependence on LAO thickness can be explained by the 
transition from ``electronic 
reconstruction'' to ``chemical 
reconstruction'' involving 
surface off-stoichiometry 
with increasing thickness of LAO capping layers, but 
that on LVO thickness seems 
complicated by the presence of a considerable 
amount of V$^{4+}$ in the thicker layers. 

The authors would like to thank A. Tanaka, M. Takizawa and 
A. Fujimori for informative discussions. 
H. W. acknowledges financial support from 
the Japan Society for the Promotion of 
Science.  This research was made possible with 
financial support from the Canadian funding organizations 
NSERC, CFI, and CIFAR. 

\bibliography{LVO1tex}

\begin{thebibliography}{10}
\bibitem{foot1}
Here we should note that there are three types of reconstruction. 
Electronic reconstruction involves a rearrangement of electrons 
keeping the total number of electrons the same. 
Ionic reconstruction involves a rearrangement of ions at the surface 
or interfaces forming a new structure which does not have 
a polarity discontinuity. 
Chemical reconstruction involves a change in the density of 
ions at the surface or interfaces.
\bibitem{footf}
Here we have to consider only the top surface 
because the compensating charge on the other side has to be 
there for charge neutrality but does not further 
affect our spectroscopy.
\end{thebibliography}

\clearpage

\begin{center}
\large\bf{Figure captions}
\end{center}

\begin{figure}
\begin{center}
\includegraphics[width=12cm]{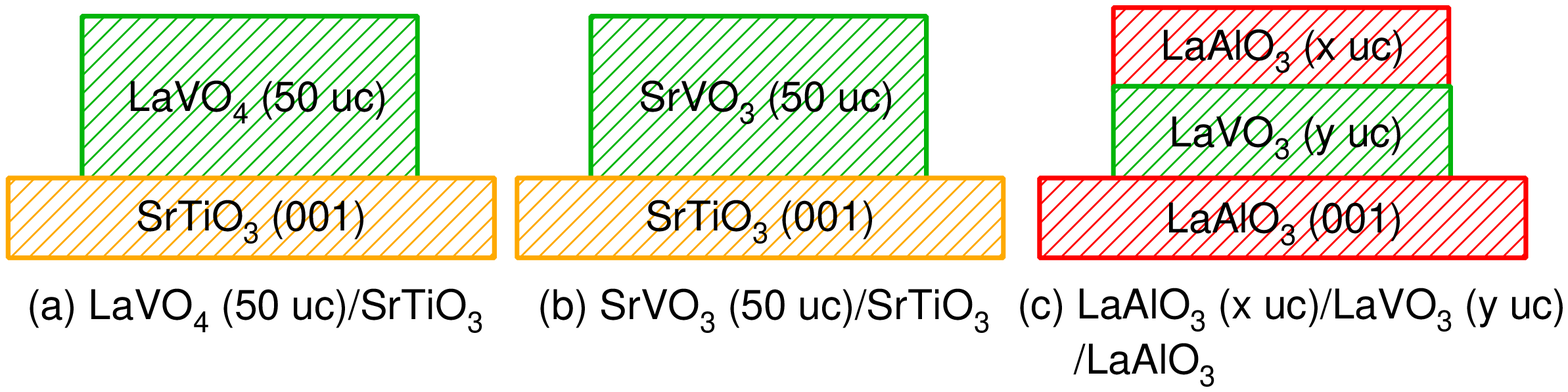}
\caption{}
\label{fig1}
\end{center}
\end{figure}

Fig.~\ref{fig1}. (Color online) 
Schematic view of the samples. 
(a) LaVO$_4$ (50 uc)/SrTiO$_3$. 
(b) SrVO$_3$ (50 uc)/SrTiO$_3$. 
(c) LaAlO$_3$ ($x$ uc)/LaVO$_3$ ($y$ uc)/LaAlO$_3$. 

\begin{figure}
\begin{center}
\includegraphics[width=14cm]{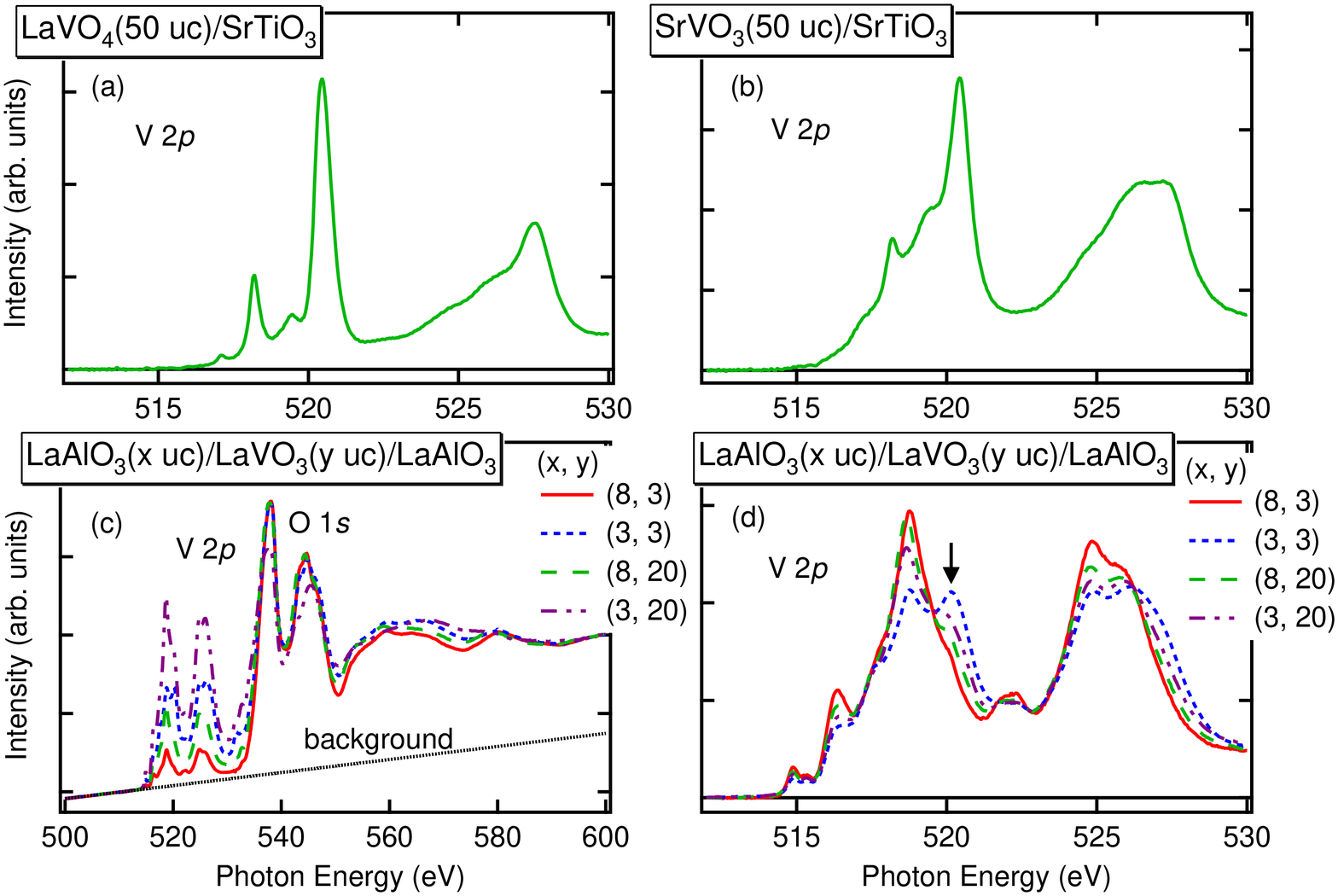}
\caption{}
\label{fig2}
\end{center}
\end{figure}

Fig.~\ref{fig2}. (Color online) 
V $2p$ XAS spectra. 
(a) LaVO$_4$ (50 uc)/SrTiO$_3$. 
(b) SrVO$_3$ (50 uc)/SrTiO$_3$. 
(c) LaAlO$_3$ ($x$ uc)/LaVO$_3$ ($y$ uc)/LaAlO$_3$ 
in a wide energy region. 
(d) LaAlO$_3$ ($x$ uc)/LaVO$_3$ ($y$ uc)/LaAlO$_3$ 
after subtracting backgrounds. 

\begin{figure}
\begin{center}
\includegraphics[width=14cm]{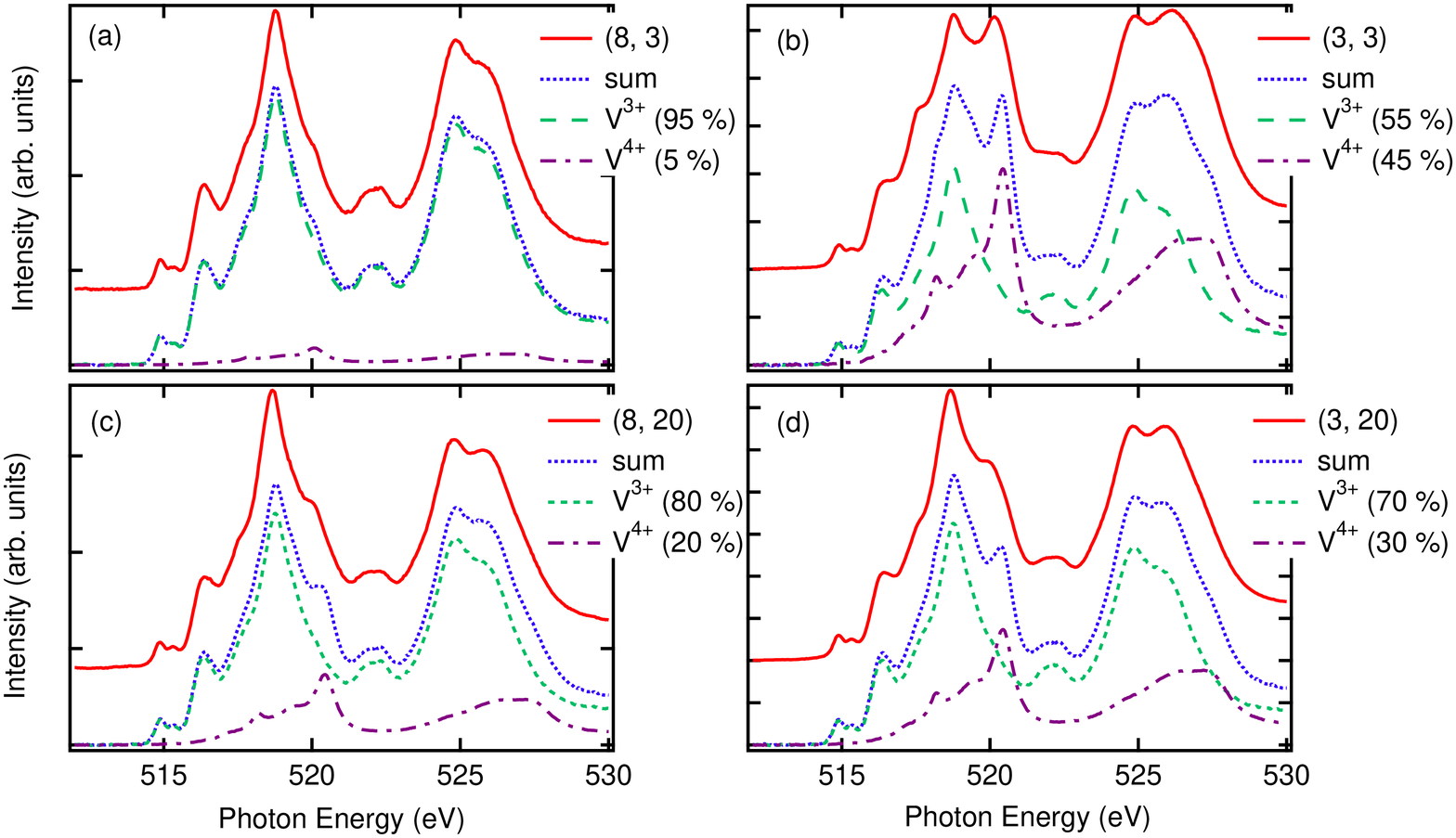}
\caption{}
\label{fig3}
\end{center}
\end{figure}

Fig.~\ref{fig3}. (Color online) 
Determination of the valence of V by using a 
linear combination of V$^{3+}$ and V$^{4+}$. 
(a) (8, 3), (b) (3, 3), (c) (8, 20), (d) (3, 20). 

\begin{figure}
\begin{center}
\includegraphics[width=16cm]{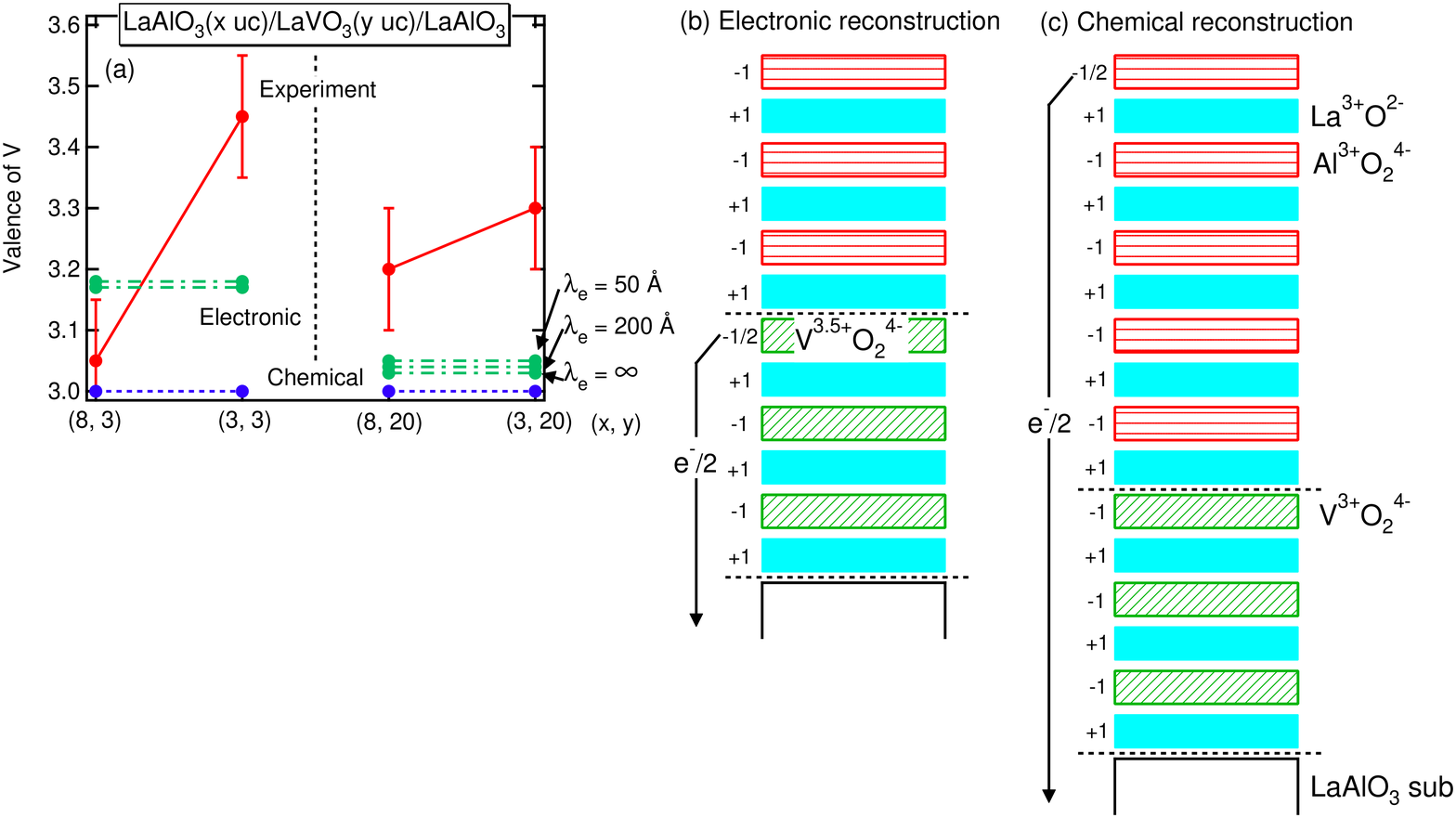}
\caption{}
\label{fig4}
\end{center}
\end{figure}

Fig.~\ref{fig4}. (Color online) 
Comparison of the valence of V between experiment 
and the reconstruction model. 
Panel (a) shows the comparison between experiment and the model, 
and panels (b) and (c) show the electronic and chemical 
reconstruction, respectively. 

\clearpage

\clearpage

\clearpage

\clearpage

\end{document}